\documentclass[
    ,final            
  ]
  {aipproc}
  
\pdfoutput=1 
\layoutstyle{8x11double}
\usepackage{natbib}

\begin{document}

\title{The Atacama B-Mode Search: CMB Polarimetry with Transition-Edge-Sensor Bolometers}


\author{T. Essinger-Hileman}{
  address={Joseph Henry Laboratories of Physics, Jadwin Hall, Princeton University, Princeton, NJ 08544, USA}
}

\author{J. W. Appel}{
  address={Joseph Henry Laboratories of Physics, Jadwin Hall, Princeton University, Princeton, NJ 08544, USA}
}

\author{J. A. Beall}{
  address={NIST Quantum Devices Group, 325 Broadway Mailcode 817.03, Boulder, CO 80309, USA}
}

\author{H. M. Cho}{
  address={NIST Quantum Devices Group, 325 Broadway Mailcode 817.03, Boulder, CO 80309, USA}
}

\author{J. Fowler}{
  address={Joseph Henry Laboratories of Physics, Jadwin Hall, Princeton University, Princeton, NJ 08544, USA}
}

\author{M. Halpern}{
  address={Department of Physics and Astronomy, University of British Columbia, Vancouver, BC, V6T 1Z1, Canada}
}

\author{M. Hasselfield}{
  address={Department of Physics and Astronomy, University of British Columbia, Vancouver, BC, V6T 1Z1, Canada}
}

\author{K. D. Irwin}{
  address={NIST Quantum Devices Group, 325 Broadway Mailcode 817.03, Boulder, CO 80309, USA}
}

\author{T. A. Marriage}{
  address={Department of Astrophysical Sciences, Peyton Hall, Princeton University, Princeton, NJ 08544, USA}
}
  
\author{M. D. Niemack}{
  address={NIST Quantum Devices Group, 325 Broadway Mailcode 817.03, Boulder, CO 80309, USA}
}  

\author{L. Page}{
  address={Joseph Henry Laboratories of Physics, Jadwin Hall, Princeton University, Princeton, NJ 08544, USA}
}

\author{L. P. Parker}{
  address={Joseph Henry Laboratories of Physics, Jadwin Hall, Princeton University, Princeton, NJ 08544, USA}
}

\author{S. Pufu}{
  address={Joseph Henry Laboratories of Physics, Jadwin Hall, Princeton University, Princeton, NJ 08544, USA}
}

\author{S. T. Staggs}{
  address={Joseph Henry Laboratories of Physics, Jadwin Hall, Princeton University, Princeton, NJ 08544, USA}
}

\author{O. Stryzak}{
  address={Joseph Henry Laboratories of Physics, Jadwin Hall, Princeton University, Princeton, NJ 08544, USA}
}

\author{C. Visnjic}{
  address={Joseph Henry Laboratories of Physics, Jadwin Hall, Princeton University, Princeton, NJ 08544, USA}
}

\author{K. W. Yoon}{
  address={NIST Quantum Devices Group, 325 Broadway Mailcode 817.03, Boulder, CO 80309, USA}
}

\author{Y. Zhao}{
  address={Joseph Henry Laboratories of Physics, Jadwin Hall, Princeton University, Princeton, NJ 08544, USA}
}

\begin{abstract}
The Atacama B-mode Search (ABS) experiment is a 145 GHz polarimeter designed to measure the B-mode polarization of the Cosmic Microwave Background (CMB) at large angular scales. The ABS instrument will ship to the Atacama Desert of Chile fully tested and ready to observe in 2010. ABS will image large-angular-scale CMB polarization anisotropies onto a focal plane of 240 feedhorn-coupled, transition-edge sensor (TES) polarimeters, using a cryogenic crossed-Dragone design. The ABS detectors, which are fabricated at NIST, use orthomode transducers to couple orthogonal polarizations of incoming radiation onto separate TES bolometers. The incoming radiation is modulated by an ambient-temperature half-wave plate in front of the vacuum window at an aperture stop. Preliminary detector characterization indicates that the ABS detectors can achieve a sensitivity of 300 $\mu K \sqrt{s}$ in the field. This paper describes the ABS optical design and detector readout scheme, including feedhorn design and performance, magnetic shielding, focal plane architecture, and cryogenic electronics.
\end{abstract}

\maketitle


\section{Introduction}

The Atacama B-Mode Search (ABS) aims to probe the physics of the early universe through measurements of the CMB polarization anisotropies. Models of inflation predict that a gravitational-wave background existed in the early universe. Such a background would leave its imprint on the CMB in the form of a pseudo-scalar B-mode component in the CMB polarization anisotropies, whereas most other sources of CMB polarization only create vector-like E-modes \cite{b-modes1, b-modes2}. Detecting a B-mode component of the CMB anisotropies is a primary goal of a number of current and upcoming experiments. The ABS experiment is designed for rapid deployment to a high-altitude site in the Atacama Desert of Chile. The entire experiment will be assembled inside a modified shipping container in North America and shipped to Chile ready to rise out of the container roof and observe soon after it arrives. 

\begin{table}
\begin{tabular}{lll}
\hline
  \tablehead{1}{l}{b}{Parameter}
  & \tablehead{1}{l}{b}{Value}
  & \tablehead{1}{l}{b}{Units} \\
\hline
Angular Resolution & 35 & Arcminutes \\
Frequency Coverage & 127-160 & GHz \\
Sky Coverage & 500 & Square Degrees \\
Multipole Coverage & 25 - 200 &  \\
Pol. Modulation & Warm HWP &  \\
Location & Ground (Chile) &  \\
Instrument NEQ & 15 & $\mu K$ $s^{1/2}$ \tablenote{Calculated for 240 detectors each with two bolometers of sensitivity 300 $\mu$K s$^{1/2}$. This sensitivity is extrapolated from dark tests of prototype detectors.} \\
\hline
\end{tabular}
\caption{Experimental Parameters for ABS}
\label{parameters}
\end{table}

\begin{figure}
  \includegraphics[height=.4\textheight]{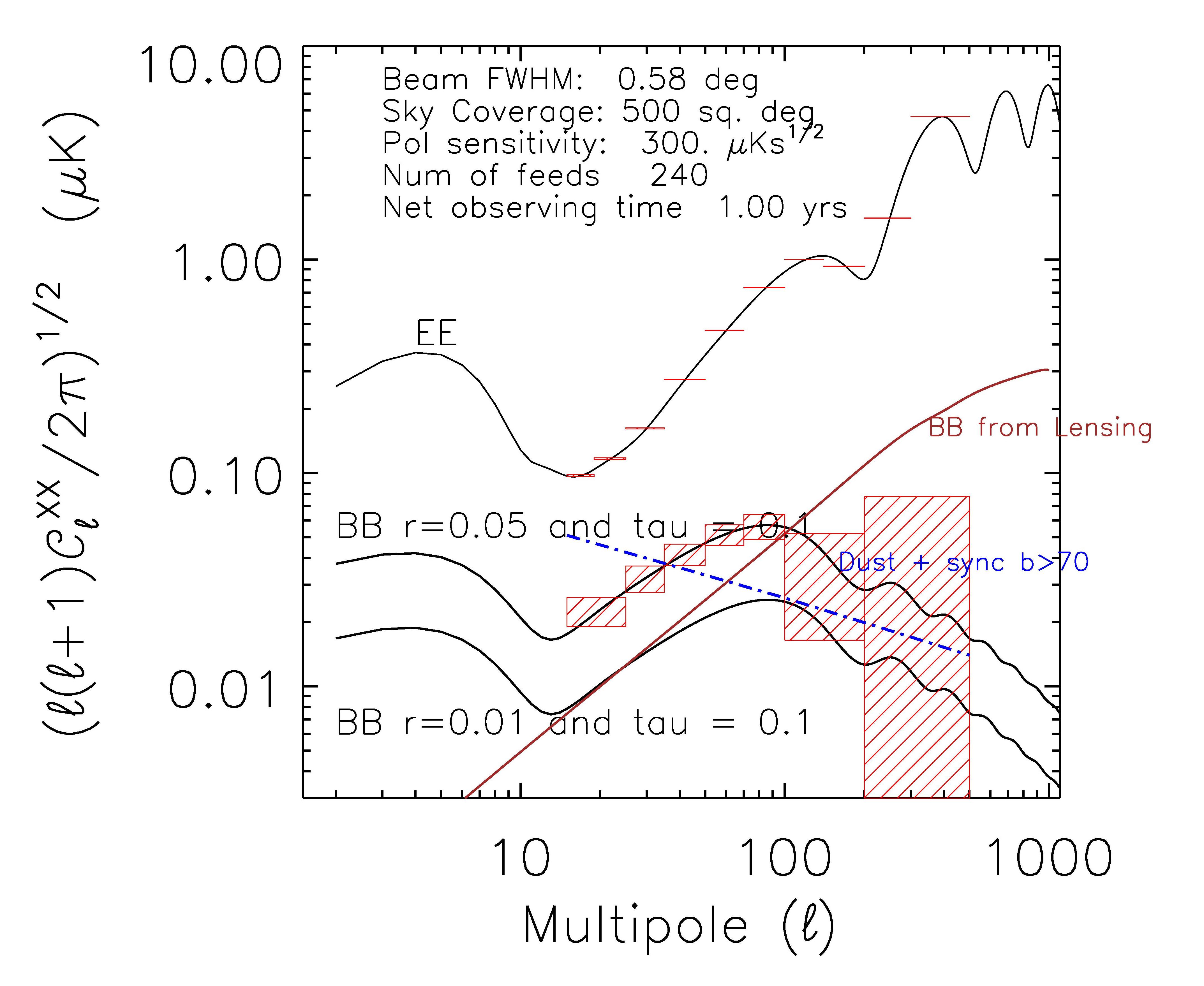}
  \caption{Projected sensitivity of ABS to the EE and BB power spectra. The top curve is a model EE power spectrum for a $\Lambda$CDM cosmology with parameters currently favored by WMAP \cite{dunkley2}. The bottom two solid black curves are the projected BB power spectra for tensor-to-scalar ratios $r=0.05$ and $r=0.01$ and optical depth $\tau=0.1$. Projected foregrounds include polarized galactic dust (blue curve), estimated from \cite{dunkley1}, and B-modes from lensing (red curve). Estimated binned errors for the EE spectrum and the BB spectrum with $r=0.05$ are shown as hashed red boxes.}
  \label{sensitivity}
\end{figure}

ABS will observe two 250-square-degree patches around the south galactic pole with $35^{'}$ beams spread over a $24^{\circ}$ instantaneous field of view. To reduce systematic errors, a warm half-wave plate (HWP) will modulate the incoming polarization before the radiation enters the beam-forming optics, allowing for better discrimination of atmospheric and instrumental noise sources. ABS is projected to be most sensitive to B-modes in the $\ell=80-120$ range, as shown in Figure \ref{sensitivity}.

\section{Receiver}

\begin{figure}
\includegraphics[height=.35\textheight]{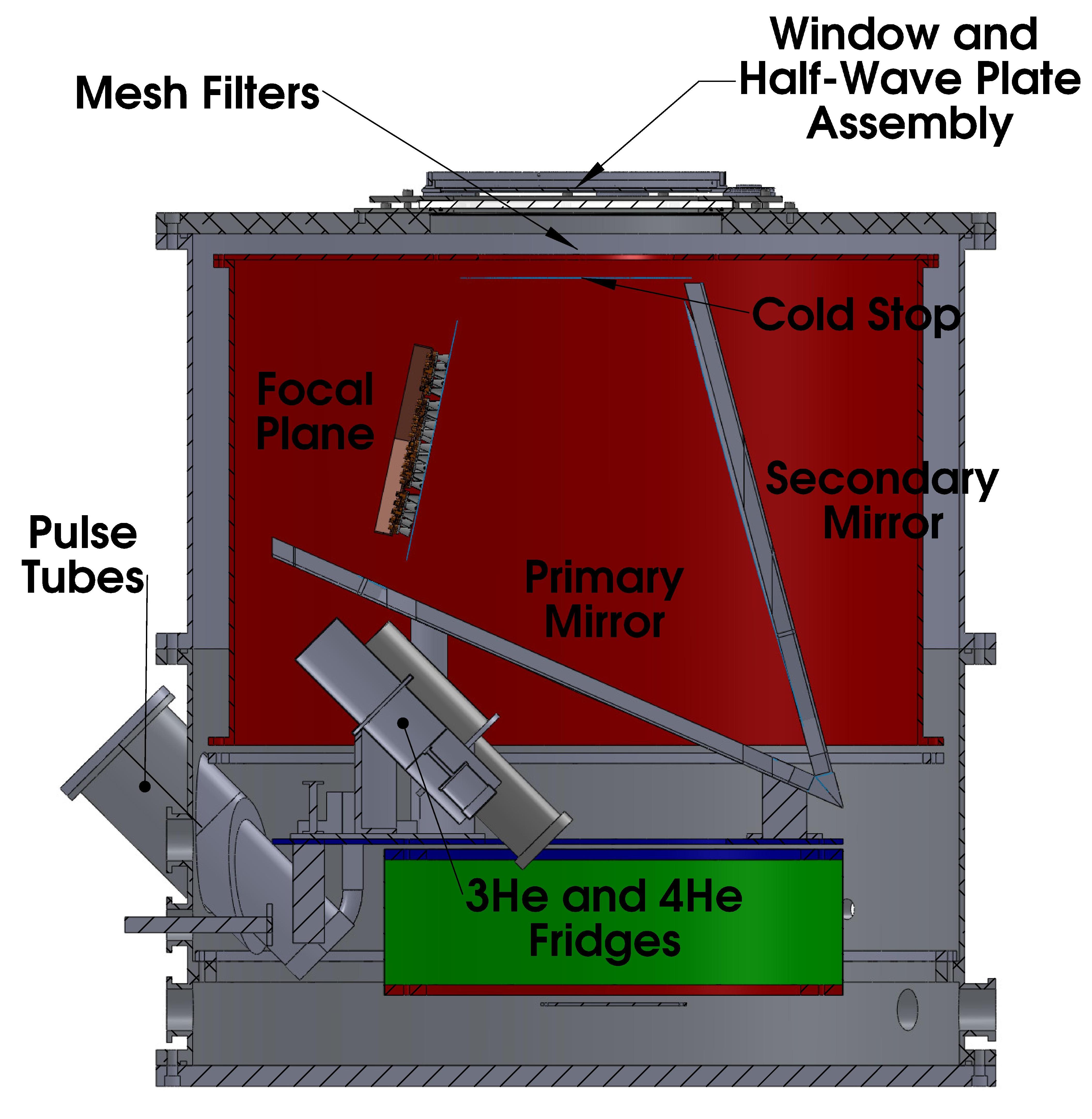}
\caption{Layout of the ABS receiver, showing the positions of the pulse tubes, the $^{3}$He and $^{4}$He absorption fridges, and the major optical elements. Light from the sky enters from the top of the figure. During observations, the receiver will be tilted at approximately 45$^{\circ}$ so that the pulse tubes are vertical.}
\label{dewar}
\end{figure}

The ABS receiver is shown in Figure \ref{dewar}. The detectors need to be cooled to 300 mK to operate. In addition, ABS will have its primary and secondary mirrors cooled to 4 K to reduce thermal loading on the detectors, provide a cold and stable surface for beam spillover, and eliminate cryogenic lenses. To achieve this, the ABS cryostat is cooled by two pulse tube cryocoolers, each of which has a 40 K and a 4 K cold head. Cryogenic stages at 1 K and 300 mK are provided by $^{4}$He and $^{3}$He absorption fridges, respectively. In addition, the cryogenic system for ABS was designed to accommodate two levels of magnetic shielding in addition to the shielding directly around the detectors at the focal plane. A mu-metal shield at room temperature and a Cryoperm shield at 4 K should provide a shielding factor of at least 100.

\section{Optics}

\begin{figure}
\includegraphics[height=.33\textheight]{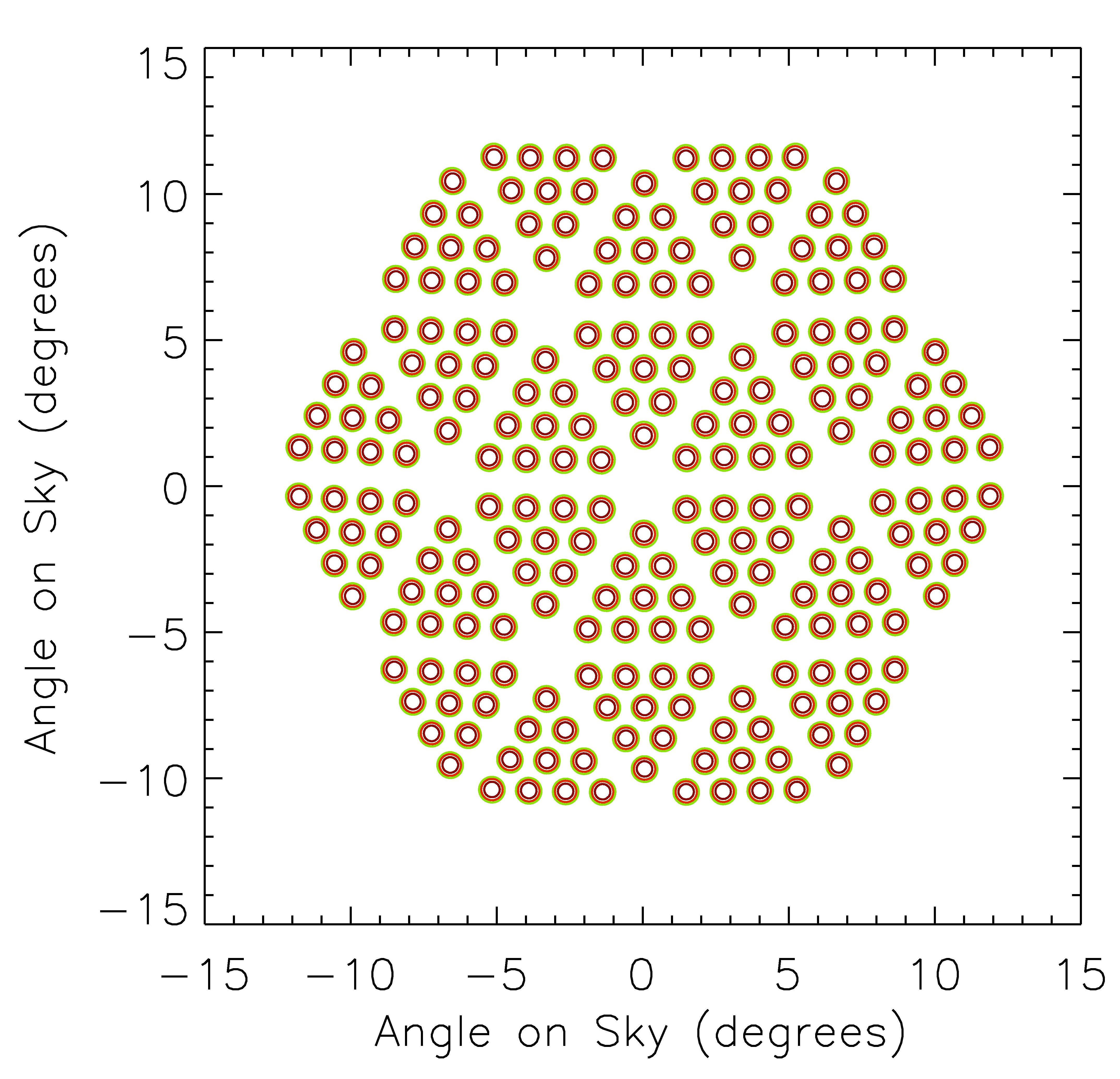}
\caption{Simulated ABS beams for one polarization at 145 GHz from DADRA. The contours are at -3, -6, and -9 dB for each of the 240 detectors. Axes are in degrees. Beams have full-width half maxima of $35^{\prime}$.}
\label{beams}
\end{figure}

ABS will use 60-cm mirrors in a compact crossed-Dragone configuration \cite{dragone1, dragone2}, which allows for both mirrors to be kept inside the cryostat at a temperature of 4 K. In addition, the crossed Dragone design was optimized to give good focus and low cross polarization over a large focal plane. The optical design was initially optimized using CodeV, a geometric ray-tracing software package. Further analysis was carried out using DADRA \cite{dadra}, which performs full numerical calculations including diffraction. The simulated beams on the sky in the center of the ABS passband at 145 GHz from DADRA are shown in Figure \ref{beams}.

In order to accomodate a focal plane of 240 feedhorn-coupled detectors, large throughput input optics are required. A vacuum window 330 mm in diameter will be made of ultra-high molecular weight polyethylene (UHMWPE) anti-reflection coated with expanded PTFE. A warm sapphire half-wave plate (HWP), also 330 mm in diameter, will be rotated in front of the vacuum window to modulate the polarization of incoming radiation. Large-format metal-mesh filters in the optical path will reduce loading on the cryogenic stages.

\section{Focal Plane Layout}
The 240 ABS detectors are grouped into triangular pods bolted into a copper support structure that mechanically supports the pods and conducts heat away. Each ABS pod consists of ten feedhorns that are supported by an aluminum interface plate. The interface plate sets the polarization angle, which is unique for each pod, of each feedhorn and forms part of an overall superconducting magnetic shield, along with an aluminum lid on the back of the pod. The detector chips are glued on the back of the feedhorns, at the output of a circular section of waveguide. Wirebonds connect the detector chips to a custom flexible/rigid readout circuit that carries the signals to microfabricated chips which house the shunt resistors, Nyquist inductors, and the first two stages of multiplexing Superconducting QUantum Interference Device (SQUID) amplifiers for the pod. The readout circuit has two rigid triangular pieces bridged by a short flexible section, which allows the circuit to fold over itself in a compact arrangement. A further flexible section on the circuit carries the critical lines from the pod, out a thin slot in the aluminum lid, and to zero insertion force connectors on another board. From there superconducting wires carry the signals to a SQUID series array for amplification at 4 K and then on to room temperature. Additional magnetic shielding at 300 mK is provided by 0.5-mm-thick niobium sheets above and below the board containing the SQUID amplifiers.

\begin{figure}
\includegraphics[height=.26\textheight]{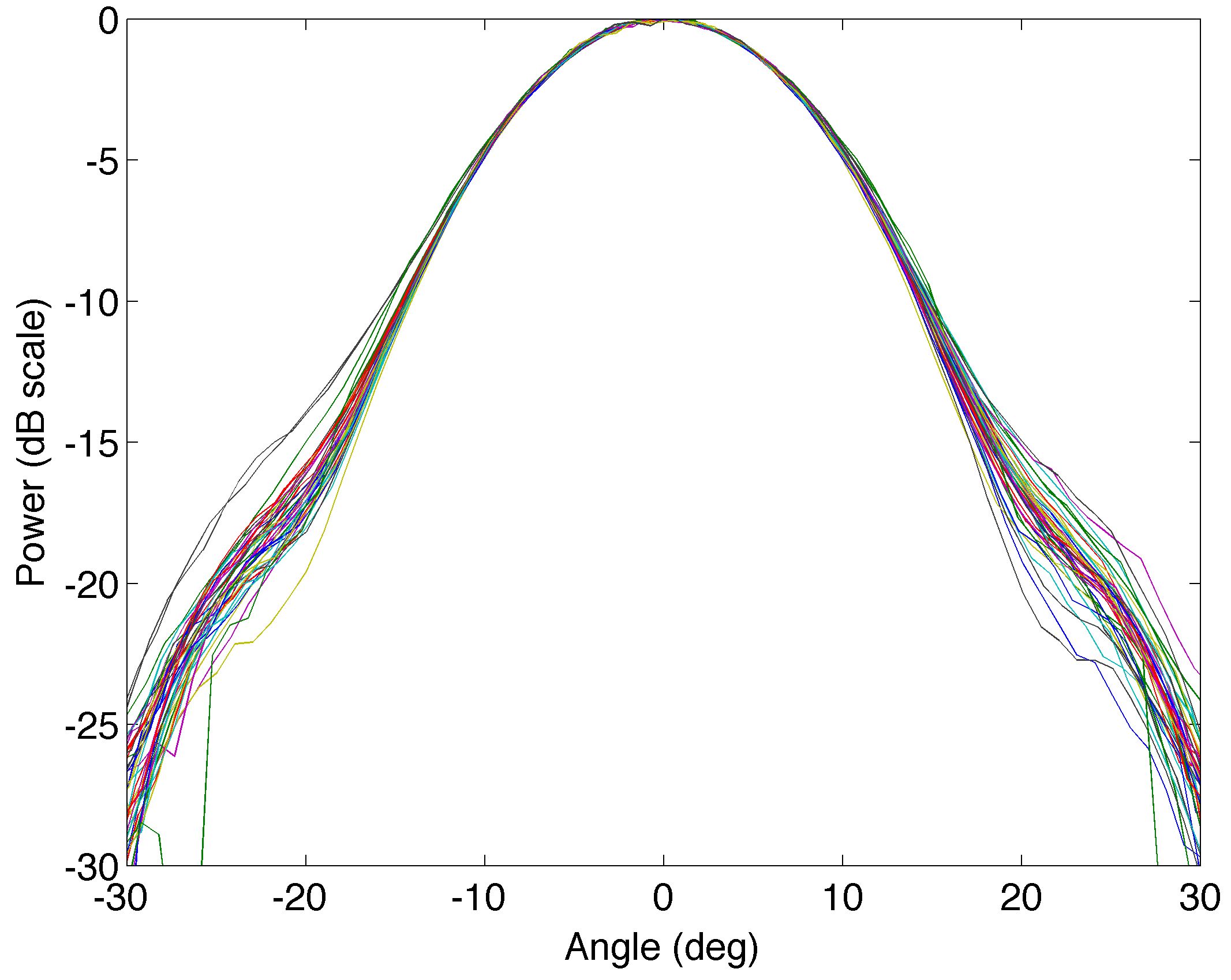}
\caption{Measured H-plane beam maps for the first 42 machined aluminum feedhorns. Vertical axis is dB from maximum. Horizontal axis is in degrees from the maximum position.}
\label{beam_maps}
\end{figure}

The feedhorns for ABS were designed with aid from simulations in the Conical Corrugated Horn Analysis (CCORHRN) program. Because of the small dimensions involved, machining restrictions required that the feedhorns be machined in two pieces. A beam-testing setup has been built to test each feedhorn assembly individually to make sure that it is free of chips or defects. A composite beam map from the first 42 feedhorns is shown in Figure \ref{beam_maps}. The beams match theoretical predictions at the level of a few percent.

\section{detectors}

\begin{figure}
  \includegraphics[height=.3\textheight]{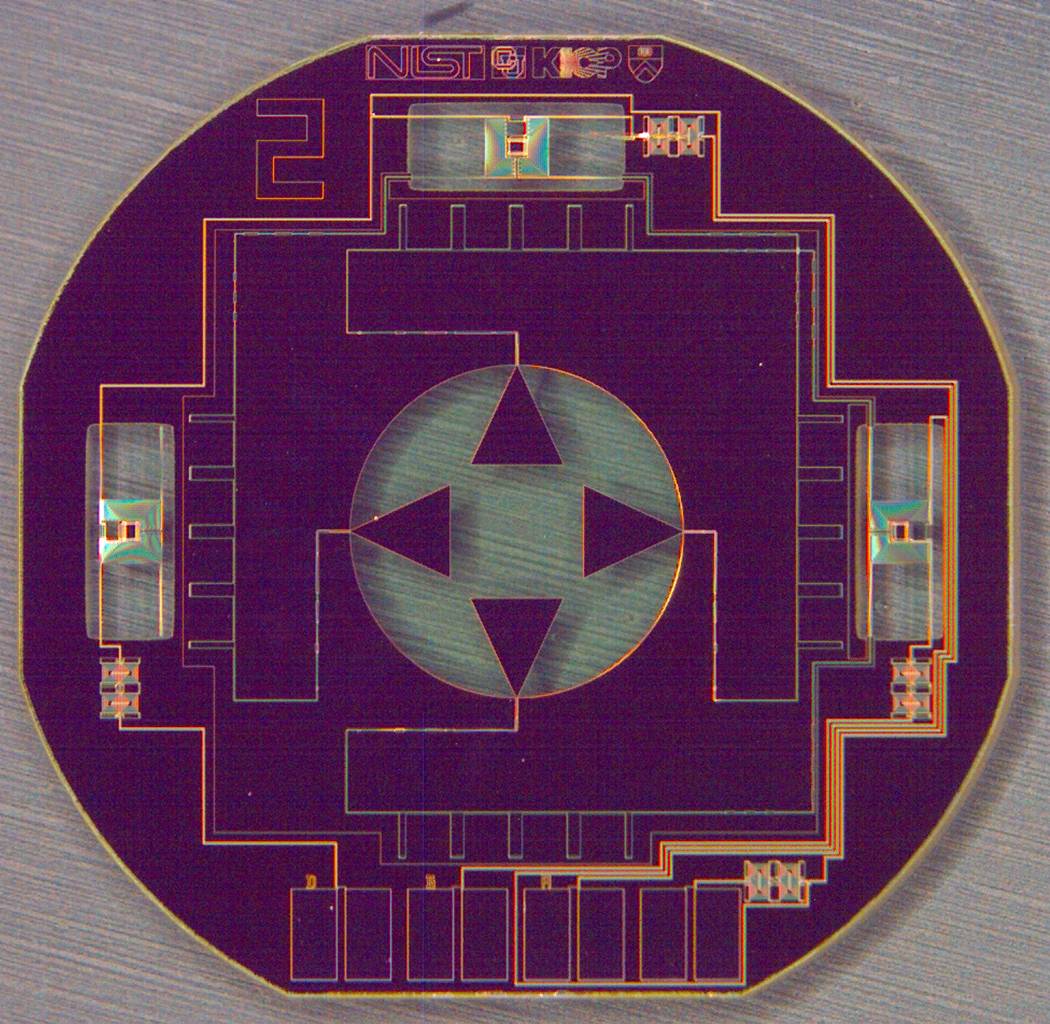}
  \caption{Photograph of a completed detector chip. The triangular leads of the OMT are visible suspended on the clear silicon nitride membrane in the center of the chip. Three TES islands are visible at the left, top, and right of the figure, one of which is not coupled optically and will be used for calibration. Bond pads for connecting to the readout circuit are at the bottom of the figure.}
  \label{chip}
\end{figure}

The ABS detectors are fabricated at NIST and were designed and tested as part of a collaboration between NIST, Princeton University, the University of Chicago, and the University of Colorado at Boulder. As shown in Figure \ref{chip}, each detector consists of a planar ortho-mode transducer (OMT) that couples the two orthogonal polarizations of light onto separate microstrip lines. The OMT, which has triangular niobium probes suspended on a thin silicon nitride membrane, couples incoming radiation from a corrugated feedhorn through a coplanar-waveguide-to-microstrip transition. After passing through on-chip lowpass filters, the microstrip lines terminate in lossy meanders that deposit power onto separate TES bolometers. For more details on detector design and characterization see \cite{appel, austermann, bleem, mcmahon, yoon}. Extrapolating data taken from dark tests and assuming a 50\% optical efficiency, the ABS detectors should achieve a sensitivity of 300 $\mu K \sqrt{s}$ in the field.

The TES bolometers will be time-domain multiplexed and amplified through three stages of SQUIDs, read out by a Multi-Channel Electronics (MCE) system provided by the University of British Columbia. This will allow for the 240 polarimeters in ABS to be read out with a minimum of wiring to room temperature, which is important in reducing thermal loading on the 300 mK stage of the cryogenics.

\section{Conclusions}
The ABS design and construction is underway and on schedule for deployment in 2010.

\begin{theacknowledgments}
  
We would like to thank our collaborators in the detector development at the Kavli Institute for Cosmological Physics and the University of Colorado at Boulder. We also gratefully acknowledge Bill Dix and Glenn Atkinson for skillfully machining many parts of the experiment, notably the mirrors and feedhorns, as well as Jennifer Lin, who designed the feedhorns. Work at NIST is supported by the NIST Innovations in Measurement Science program. Work at Princeton University is supported by the NSF through awards PHY-0355328 and PHY-085587 and NASA through award NNX08AE03G. T. Essinger-Hileman was supported by a National Defense Science and Engineering Graduate (NDSEG) Fellowship.

\end{theacknowledgments}




\end{document}